# The *inferred* redshift distribution of the Faint Blue Galaxy Excess[1]

Simon P. Driver and Warrick J. Couch

School of Physics, University of New South Wales, Sydney, NSW 2052, Australia

Steven Phillipps

Astrophysics Group, Department of Physics, University of Bristol, Bristol BS8 1TL, UK

and

Rogier A. Windhorst

Department of Physics and Astronomy, Arizona State University, Box 871504, Tempe, AZ 85287-1504, USA

astro-ph/9605048  10 May 1996

## ABSTRACT

We infer the redshift distribution of the faint blue galaxy excess (FBE) at $m_B = 23.5$ by subtracting the predicted distribution of giant/normal galaxies from the observed N($z$) distribution for all types. This is possible because of the recent deep *Hubble Space Telescope* (HST) WFPC2 morphological number counts which have convincingly demonstrated that little evolution of the giant population is seen to $m_B = 26.0$. The mean redshift of the FBE at $m_B = 23.5$ is found to be $<z>_{FBE} = 0.40 \pm 0.07$ with upper and lower quartiles defined by $z_{0.75} = 0.58 \pm 0.05$ and $z_{0.25} = 0.28 \pm 0.05$, respectively.

We compare this inferred FBE N($z$) distribution to the predictions from three generic faint galaxy models: dwarf dominated (no evolution), pure luminosity evolution, and evolving dwarfs. The inferred FBE N($z$) distribution strongly supports a hybrid evolving dwarf–rich model wherein a large population of dwarfs present at z=0.5 has subsequently faded to obscurity. The total integrated number density of dwarfs (down to $M_B = -11$) is estimated to be a factor of 20 times greater than that of E—Sc galaxies and the estimated fading to be $1.0 < \Delta m < 1.4$ mags. Thus, the dwarf population is estimated to be responsible for $\sim 30\%$ of the luminosity density locally, rising to $\sim 57\%$ at $z = 0.5$.

*Subject headings:* galaxies: elliptical — galaxies: spiral — galaxies: irregular — galaxies: evolution — galaxies: formation

## 1. INTRODUCTION

Morphological number counts for E/S0 and Sabc galaxies from the HST Medium Deep and Ultra Deep WFPC1 and WFPC2 surveys are well fit by models with only passive or weak evolution to at least $B = 26$ [Casertano *et al.* 1995; Glazebrook *et al.* 1995 (hereafter G95); Driver, Windhorst & Griffiths 1995a (hereafter DWG); Driver *et al.* 1995b, 1996; Abraham *et al.* 1996a,b]. The observed morphological number counts of late–type spirals, however, are an order of magnitude greater than the prediction based on the standard no–evolution Sd/Irregular models (Driver *et al.* 1995b). These galaxies thus contribute strongly to the well known faint blue galaxy excess (FBE).

Correspondingly, the Canada-France-Redshift-Survey (CFRS) has measured the redshifts for $\sim 591$ galaxies in the magnitude range $17.5 < I_{AB} < 22.5$ with a completeness of $\sim 75\%$ (Lilly *et al.* 1995). Separating the sample according to color, these authors find no evidence for a change in the field luminosity function (LF) for the red (Sbc and earlier) population over a wide range of redshift ($0.3 < z < 1.0$). However, the LF for the blue population (Sc and later) shows a significant steepening and brightening with increasing redshift. Similarly, Colless (1995) and Ellis *et al.* (1996) show that although a change in the overall LF is seen with redshift, it occurs predominantly amongst the lower luminosity population, dominated by later-types and dwarf systems. Steidel, Dickinson & Persson (1994) also conclude that there has been little evolution in the giant galaxy population out to $z \sim 1$ by the complimentary route of studying the systems responsible for MgII quasar absorption lines.

While some cosmic conspiracy cannot be ruled out, for example early-type systems are being formed and destroyed at the same rate, the simplest interpretation is that such systems have evolved slowly from $z \sim 1$ to the present day. Given this conclusion and the recent measures of the local LF for giant galaxies (e.g. Marzke *et al.* 1994), it becomes a relatively trivial exercise to compute the predicted N($z$) distribution for giant galaxies (which we formally define to be E–Sc Hubble types) at any magnitude interval down to $B = 26$. However, two questions do remain: normalisation of the LF and the choice of cosmological model. Arguments put forward by DWG and G95 support normalization of faint galaxy models to the observations at $B = 18$ (see also Driver & Phillipps 1996) and we shall adopt their philosophy here. With regards the choice of world models, Driver *et al.* (1996) discuss constraints on the cosmological parameters from the HST elliptical galaxy number counts alone, and conclude that a high-$\Lambda$ ($\Omega_\Lambda > 0.8$) universe can be ruled out and that a $\Omega = 1, \Lambda = 0$ model (with passive evolution) provides the optimal fit to the data. Taking this as reasonably established leads us to the conclusion that the N($z$) for E/S0 and Sabc systems can be predicted with confidence. Hence the N($z$) distribution of the FBE (assumed to be late–type dwarfs) can simply be inferred by subtracting the giant N($z$) from the observed total N($z$); we derive this distribution in §2 and compare it to three generic faint galaxy models in §3. We summarize the results in §4.

## 2. THE INFERRED FAINT BLUE GALAXY REDSHIFT DISTRIBUTION

The most recent and faintest redshift surveys available are those of G95 and Cowie, Songaila & Hu (1991, hereafter CSH). G95 obtained redshifts for a sample of $\sim$ 90 galaxies in the magnitude interval $22.5 \leq B_J \leq 24$ were with $\sim$ 80% completeness, while CSH obtained redshifts for 11 of 12 galaxies in the same range. The combined data set, although of limited size, represents a near complete sample representative of the overall field galaxy population at moderately faint magnitudes, where the number counts are observed to be 2–3 times higher than can be accounted for by standard no–evolution models (e.g. Broadhurst, Ellis & Shanks 1988).

It has been shown repeatedly (see Koo & Kron 1992 and references therein) that invoking evolution of the whole population is unable to match the observed redshift distributions. The recent additional evidence of the limited evolution of giant galaxies via morphological studies is perhaps the single most important breakthrough in this topic. It firstly confirms that the number counts, at least for specific morphological types, can be well modelled and secondly allows the population responsible for the FBE to be isolated.

Figure 1 shows the observed redshift distribution (histogram) taken from G95 for 72 out of 91 galaxies in the magnitude interval $22.5 \leq B_J \leq 24$. The shaded regions in the upper panels are the predicted giant N($z$) distributions based on either the Mt Stromlo/APM (Loveday *et al* 1992; hereafter SAPM) or the CFA1&2 (Marzke *et al.* 1994; hereafter CFA12) LFs. The exact LF parameters used are those tabulated in DWG with K-corrections and modelling as in Driver *et al.* (1994). These models are normalized at $B_J = 18$ as discussed above. Mild luminosity evolution (equivalent to the passive evolution of old stellar populations; Bruzual & Charlot 1993; see also Pahre, Djorgovski & de Carvalho 1996, Im *et al.* 1996) is adopted in these models and is equivalent to a 0.4 magnitude brightening by $z = 0.5$ (see Driver *et al.* 1996). (If a no–evolution model were adopted instead, it would move the peak of the predicted giant N($z$) distribution by less than $\Delta z = 0.04$ and decrease the amplitude by $\sim$ 10%.)

The lower panels of Figure 1 show the resultant FBE N($z$) distributions based on the subtraction of the mild-evolution giant N($z$) models from the observed N($z$) distribution. Given the sparsity of the data, we parameterize the inferred FBE N($z$) distributions simply by their mean and quartiles, ie. $<z>_{SAPM}$

$= 0.37^{0.50}_{0.27}$, or $<z>_{CFA12} = 0.44^{0.65}_{0.29}$. The equivalent values for the original sample of all types are $<z> = 0.43^{0.59}_{0.29}$. It can be seen that the results are significantly different depending upon whether the derived giant LF from the SAPM survey or the CFA12 survey is used; this is a cause for considerable concern. Examining the two sets of input LF parameters, it is possible to see why this difference occurs. The SAPM survey has a significantly brighter $M^*$ which produces a substantially larger $<z>$ for the giants. When subtracted from the observed N($z$) distribution, the resulting distribution therefore has significantly fewer high–z objects than for the CFA12 survey, and the distribution is skewed towards lower redshift. Given that the two surveys quote errors which exclude each other's results, we shall simply adopt the range of the two predictions as an empirical estimate of the error in the FBE mean redshift. We conclude that **the mean redshift of the FBE at $m_B = 23.5$ lies between 0.37 and 0.44.**

## 3. POPULAR GENERIC FAINT GALAXY MODELS

We now attempt to reproduce the inferred FBE N($z$) by considering three generic faint galaxy models: a no–evolution dwarf–rich model (eg. Driver *et al.* 1994), a pure luminosity evolution model (eg. Metcalfe *et al.* 1991), and a hybrid mixture of the two based on an evolving dwarf–rich population (eg. Phillipps & Driver 1995; hereafter PD95). While this is not an exhaustive comparison – for example we do not consider the merging model of Broadhurst, Ellis & Glazebrook (1992; but see G95 for a comprehensive discussion) – these models have, in combination, been used with considerable success in fitting the HST morphological number counts (DWG, Driver *et al.* 1995b).

The dwarf–rich models exploit the large uncertainty in the local space density of dwarfs (see Schade & Ferguson 1994, Driver & Phillipps 1996) and increase the normalization and/or slope of the faint end of the LF until an optimal fit is found (eg. Koo, Gronwall & Bruzual 1993). Supporting evidence comes from the similar blue colors of late–type dwarfs and the FBE and the small half-light radii of the FBE reported by Im *et al.* (1995). Dwarf–rich, no–evolution models *can* provide good fits to the counts in all bands as well as naturally explain the trend to bluer colors at fainter magnitudes, but fail to match the overall N($z$) distribution. To illustrate this we adopt Schechter(1976) LF parameters of $M^*_{Dwarf} = -18.0$, $\alpha_{Dwarf} = -1.8$ and increase the normalization $\phi^*_{Dwarf}$ until a match to the number counts at $m_B = 23.5$ is achieved. The resulting dwarf N($z$) distribution for this magnitude interval is shown in the lower panels of Figure 1 and has $<z>_{Dwarf+NoEvol} = 0.12^{0.20}_{0.06}$.

Note that the slight difference in amplitude between the left and right lower panels is due to the differing fractional contribution of giants to the total number counts at $m_B = 23.5$. Clearly the model severely over-predicts the density of low redshift objects and the distribution is far too steep and narrow. A *non-evolving* population of low luminosity systems can therefore be ruled out (cf. PD95).

Evolving standard models assume the LF for late–type dwarfs is an extrapolation of the standard one for luminous giants (ie. $\alpha = -1$ as in SAPM), and invoke luminosity evolution at the faint end to match the counts. Adopting $\phi^*_{Dwarf} = 0.11 \, \text{Mpc}^{-3}$, $M^*_{Dwarf} = -18.5$, $\alpha_{Dwarf} = -1.1$ (based on the typical giant LF extrapolation; DWG) results in a severe under-prediction of the dwarf number counts. Hence pure luminosity evolution[2] is added until a match to the counts at $m_B = 23.5$ is made. We approximate the luminosity evolution by a linear variation in magnitude with look–back time (PD95) ie. $\Delta m = 4\beta[1 - (1+z)^{-\frac{3}{2}}]$. To fit the number counts at $m_B = 23.5$, the rate of luminosity evolution required is $\beta = 1.4$ (or $\Delta m = 2.6$ at $z = 0.5$). The resulting dwarf N($z$) distribution is shown in Fig. 1 (dotted line) and is characterized by $<z>_{Dwarf} = 0.79^{1.08}_{0.52}$. The level of luminosity evolution required to match the counts is sufficiently strong as to predict a much higher mean redshift than that inferred. In order to save such models, luminosity–dependent evolution is required such as that originally proposed by Broadhurst, Ellis & Shanks (1988).

By combining the first two models, it is clear that an optimal solution should be found (see PD95). Given the sparsity of the data and the ambiguity in the giant LFs we find the optimal combinations which match the two redshift distributions and use these to define the most likely range of parameters.

1. Consider first the N($z$) distribution based on assuming the SAPM LF for giants. As demonstrated above, simply adopting an extrapolation of the giant LF and invoking pure luminosity evolution to match the counts severely overpredicts $<z>_{Dwarf}$. However, by enriching the initial local luminosity function (which is poorly constrained anyway; Driver & Phillipps 1996) with additional dwarfs, we should require less evolution and hence reduce the predicted mean redshift of the FBE. Given the narrowness of the inferred FBE N($z$) distribution, simply steepening the LF is unable to reproduce the narrow distribution (without invoking a more complicated evolutionary scheme), since a wide range of intrinsic magnitudes will then contribute at any magnitude interval. Increasing the normalization, on the other hand, will generate a narrower distribution as the number counts will be dominated by the dwarf galaxies with intrinsic luminosities close to $M^*_{Dwarf}$ (in the same way that $M^*$ giant galaxies dominate the local number counts).

---

[2]This is defined as an increase in a galaxy's luminosity due to internal stellar and dynamical processes only, as opposed to tidally induced LE, merger driven LE or any other externally induced LE.

Therefore, in an attempt to match the dwarf N(z) distribution implied by the SAPM LF, we fix the faint end slope to that observed for giant galaxies and allow only the normalization of the dwarf LF ($\phi^*_{Dwarf}$) and the amount of evolution ($\beta$) to be free parameters. That is, we assume that the SAPM LF is correct and that the dwarfs also have a flat luminosity distribution. The optimal values found by simply trading off the level of evolution against $\phi^*_{Dwarf}$ in order to match the number counts are (with giant parameters from SAPM):

$$\begin{aligned} \phi^*_{Dwarf} &= 5 \times \phi^*_{Giant} &&= 7.0 \times 10^{-3} \text{ Mpc}^{-3} \\ M^*_{Dwarf} &= M^*_{Giant} + 2.5 \text{ mags} &&= -18.5 \\ \alpha_{Dwarf} &= \alpha_{Giant} &&= -1.1 \\ \beta_{Dwarf} &&&= 0.6 \end{aligned}$$

giving: $<z>_{EvolvingDwarf} = 0.38^{0.52}_{0.27}$ compared to $<z>_{SAPM} = 0.37^{0.50}_{0.27}$.

2. To match the number counts and the FBE N(z) distribution inferred from CFA12, we adopt the CFA12 $M^*$, and $\alpha$ values for dwarfs also, and allow only the amount of luminosity evolution ($\beta$) and $\phi^*$ to be free. Thus again we assume that the shape of the CFA12 LF is correct and that the dwarfs are uniformly evolving:

$$\begin{aligned} \phi^*_{Dwarf} &= 3.5 \times 10^{-4} \text{ Mpc}^{-3} \\ M^*_{Dwarf} &= -20.3 \\ \alpha_{Dwarf} &= -1.7 \\ \beta_{Dwarf} &= 0.75, \end{aligned}$$

giving: $<z>_{EvolvingDwarf} = 0.44^{0.70}_{0.25}$ compared to $<z>_{CFA12} = 0.44^{0.65}_{0.29}$.

Both solutions require a larger population of local dwarf galaxies than generally supposed hitherto. In the first solution the dwarfs are distributed with a flat (giant-like) luminosity distribution but with a normalization a factor of 5 higher than that of the giant galaxies. The second solution adopts a more normal (giant-like) normalization (1.5× that listed in DWG for Sd/Irrs) but of course has the much steeper faint end slope reported by CFA12. Both models require that these dwarfs (Sd/Irrs/dIs) undergo significant luminosity evolution. In particular, if we assume the two models bracket the optimal range of solutions to the FBE at $m_B = 23.5$, we can conclude the following:

(1) The integrated local space density of late–type dwarfs (down to $M_B = -11$) is a factor of $\simeq 23$ or 21 times higher than E–Sc's, for the SAPM– and CFA12–based models, respectively.

(2) At $z = 0.0$ the integrated luminosity density of Sd/Irr systems (derived by integrating the luminosity functions to $M_B = -11$) is 25%—35% of the total.

(3) At $z = 0.5$ the integrated luminosity density of late type/dwarf systems has risen to 56%—58% of the total from all galaxies.

(4) The entire local Sd/Irr dwarf population has, on average, faded by $1.0 < \Delta m < 1.4$ since $z = 0.5$.

## 4. CONCLUSIONS

We have inferred the redshift distribution of the FBE by the subtraction of the giant N($z$) distribution. We contend that the giant N($z$) distribution can be reasonably well predicted given that numerous independent observations find weak evolution in the giant populations out to $z \sim 1$. The principle caveat is that the number count models and observations are normalized at $m_B = 18.0$ as implied by the HST WFPC2 morphological number count data. By adopting the parameterization of the shape of the luminosity function for giant galaxies from both the SAPM and CFA12 surveys, we infer two estimates of the FBE N($z$) distribution giving a mean redshift for the FBE at $m_B = 23.5$ in the range 0.37–0.44. While the two inferred N($z$) distributions are visually distinct, there is sufficient agreement to rule against either a no-evolution dwarf–dominated model or a pure luminosity evolution sparse dwarf model. The most realistic alternative is the adoption of an over-density of dwarf galaxies at moderate redshifts which have recently undergone significant luminosity evolution (eg. PD95, Babul & Ferguson 1996).

We acknowledge support from the Australian Research Council (SPD, WJC), the Royal Society (SP) and HST grants GO.5985.01.95A and GO.2684.03.87A (RAW). We thank Paul Bristow for checking our models against his detailed simulations and for useful discussions. The referee is thanked for helpful comments on the original version of the paper.

## REFERENCES

Abraham, R.G., Tanvir, N.R., Santiago, B.X., Ellis, R.S., Glazebrook, K., & van den Bergh, S., 1996a, MNRAS, 279, L47

Abraham, R.G., van den Bergh, S., Glazebrook K., Ellis R.S., Santiago B.S., Surma P., & Griffiths R.E., 1996b, preprint

Babul, A., & Ferguson, H., 1996, ApJ, 458, 100

Broadhurst, T. J., Ellis, R. S., & Shanks, T., 1988, MNRAS, 235, 827

Broadhurst, T. J., Ellis, R. S., & Glazebrook, K., 1992, Nature, 355, 55

Colless, M. M., 1995, in "Wide Field Spectroscopy and the Distant Universe", eds Maddox, S.J., Aragon-Salamanca A, (World Scientific Press), p.263

Cowie, L.L., Songaila, A., & Hu, E.M., 1991, Nature, 354, 400

Driver, S. P., Phillipps, S., Davies, J. I., Morgan, I., & Disney, M.J., 1994, MNRAS, 266, 155

Driver, S. P., Windhorst, R. A., & Griffiths R. E. 1995a, ApJ, 453, 48 (DWG)

Driver, S.P., Windhorst, R. A., Ostrander, E.J., Keel W.C., Griffiths, R. E., & Ratnatunga, K.U., 1995b, ApJL, 449, L23

Driver, S.P., Windhorst, R.A., Phillipps, S., & Bristow, P.D., 1996, ApJ, 461, 525

Driver, S.P., & Phillipps, S., 1996, ApJ, in press

Ellis, R.S., Colless, M., Broadhurst, T.J., Heyl, J., & Glazebrook, K., 1996, MNRAS, in press

Glazebrook, K., Ellis, R. S., Colless, M. M., Broadhurst, T. J., Allington-Smith, J., & Tanvir, N., 1995a, MNRAS, 273, 157 (G95)

Glazebrook, K., Ellis, R. S., Santiago, B., & Griffiths, R.E., 1995b, MNRAS, 275, 19p

Im, M., Casertano, S., Griffiths, R.E., Ratnatunga, K.U., & Tyson, J.A., 1995, ApJ, 441, 494

Im, M., Griffiths, R.E., Ratnatunga, K.U., & Sarajedini, V.L., 1996, ApJ, 461, 79

Koo, D.C., Gronwall, C., & Bruzual, G.A., 1993, ApJ, 415, L21

Koo, D.C., & Kron, R.G., 1992, ARA&A, 30, 613

Lilly, S.J., Tresse, L., Hammer, F., Crampton, D., & Le Fevre, O., 1995, ApJ, 455, 108

Loveday, J., Peterson, B.A., Efstathiou, G., & Maddox, S.J., 1992, ApJ, 390, 338 (SAPM)

Marzke, R.O., Geller, M.J., Huchra, J.P., & Corwin Jr, H.G., 1994, AJ, 108, 437 (CFA12)

Metcalfe, N., Shanks, T., Fong, R., & Jones, L.R., 1991, MNRAS, 249, 498

Pahre, M.A., Djorgovski, S.G., & de Carvalho R.R., ApJ, 456, L79

Phillipps, S., Davies, J.I., & Disney, M.J., 1990, MNRAS, 242, 235

Phillipps, S., & Driver, S.P., 1995, MNRAS, 274, 832 (PD95)

Schade, D.J., & Ferguson, H.C., 1994, MNRAS, 267, 889

Schechter, P., 1976, ApJ, 203, 297

Steidel, C.C., Dickinson, M., & Persson, S.E., 1994, ApJ, 437, L75

FIGURES

Figure 1: The upper panels show the redshift distribution at $m_B = 23.5$ compared to the passive/weak evolution models based on the local observed luminosity functions from either the Mt Stromlo/APM (Fig. 1a, Loveday *et al.* 1992) or the CFA1&2 (Fig. 1b, Marzke *et al.* 1994) surveys. The lower panels are derived by subtracting the shaded regions (representing passively evolving giants only) from the overall observed distribution, this results in the *inferred* distribution for dwarf (ie. Sd/Irr) galaxies. The thick lines in the upper panels denotes the weak evolution N($z$) prediction of giant+dwarf galaxies. The three generic models in the lower panels are based on: a steep luminosity function for dwarfs coupled with moderate luminosity evolution (Evol+Dwarfs); a flat LF with strong luminosity-evolution (Evol only) and; a very steep unevolving LF (Dwarfs only). Note that the models shown all match the observed number counts at $m_B = 23.5$. The histogram data in the upper panels is from Glazebrook *et al.* (1995) and Cowie *et al.* (1991).

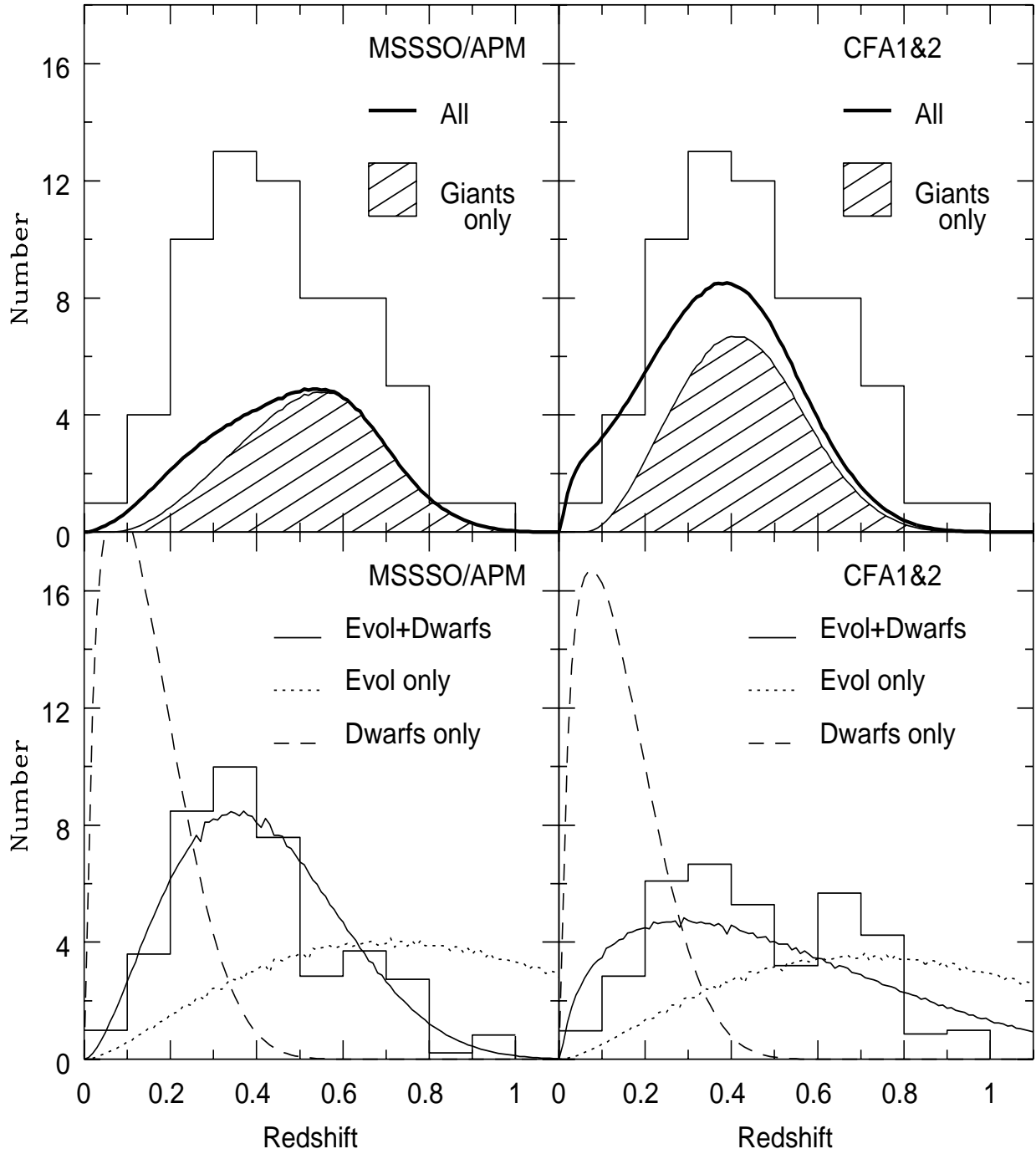